# Relaxation oscillations of Zeeman and dipole magnetizations of a paramagnet under conditions of deep low-frequency modulation

M.D. Zviadadze, A.G. Kvirikadze, R.L. Lepsveridze, G.I. Mamniashvili[*], N.M. Sozashvili

E. Andronikashvili Institute of Physics, Tamarashvili 6, 0177 Tbilisi, Georgia

[*] Corresponding author E-mail address: Mamni@iphac.ge

## Abstract

The relaxation oscillations of Zeeman and dipole magnetizaions in spin system of a solid paramagnet are theoretically analyzed under conditions of intermediate saturation of magnetic resonance and strong low-frequency modulation of the external magnetic field. Peculiarities of the relaxation oscillations in the synchronous detection regime are considered.

## 1. Introduction

Modulation of a steady magnetic field $H_0$ is commonly used to increase the signal-to-noise ratio in measurements of magnetic resonance signals [1]. To avoid the undesirable effect of the modulation on the width and shape of the resonance line, the modulation amplitude $H_M$ should be set as small as possible [2, 3].

On the other hand, the modulation itself can result in some interesting physical effects such as the enhancement of longitudinal susceptibility under conditions of shifted spin-spin temperature, as it was theoretically predicted [4] and experimentally observed by V. Atsarkin et al. [5]; the modulation saturation of the lock-in detection signals [1]; the generation of $10^2$–$10^3$ times enhanced second harmonic of the longitudinal magnetization [6], etc.

In the present article, a new modulation effect is theoretically investigated which consists in the appearance of relaxation oscillations of Zeeman and dipole magnetizations under conditions of a deep low-frequency modulation [6, 7]. It is also discussed Some peculiarities of the lock-in detection signals in the relaxation oscillation mode and possibilities for practical applications are discussed as well.

## 2. The basic equations

As it is well known [8], the spin system (SS) dynamics of solid paramagnet in the time scale of $t > T_2$ ($T_2$ is the transverse spin relaxation time) in the high-temperature approximation is described by inverse spin temperatures $\beta_z(t)$ and $\beta_d(t)$ corresponding to Zeeman and dipole subsystems. In the following it is convenient to use Zeeman $M_z(t)$ and $M_d(t)$ magnetizations:

$$M_z(t) = M_0 \frac{\beta_z(t)}{\beta_L}, \quad M_d(t) = \frac{\omega_d}{\omega_0} M_0 \frac{\beta_d}{\beta_L}, \tag{1}$$

where $M_0 = \chi_0 H_0$ is the equilibrium magnetization, $\chi_0 = \hbar^2 \gamma^2 \frac{S(S+1)}{3} \beta_L$ is the static susceptibility, $\beta_L$ is the inverse equilibrium temperature of the lattice (in energy units), $\omega_0 = \gamma H_0$ is the Zeeman frequency, $\gamma$ is the gyromagnetic ratio, $\omega_d = \{SpH_d^2 /(\hbar^2 SpS_z^2)\}^{1/2}$, where $\hbar\omega_d$ is the energy "quantum" of the dipole subsystem, and $H_d$ is the secular part of spin dipole-dipole interaction [1].

Let us insert the SS in a strong steady magnetic field $\vec{H}_0 \uparrow\uparrow z$ ($\omega_0 >> \omega_d$) and an oscillating, linearly polarized magnetic field $\vec{H}_1(t) = \vec{i} \cdot 2H_1 \cos\Omega t$ with the amplitude $2H_1$ and frequency $\Omega$ ($\vec{i}$ is the $x$ axis unit vector, $T_1^{-1} << W << T_2^{-1}$, $T_1$ is the spin-lattice (S-L) relaxation time, $W$ is the probability of the spin reorientation induced by the oscillating field $2H_1, \Omega$). Let us assume that this system is also subjected to the modulation field $H_M \cos\omega t$ parallel to $\vec{H}_0$. In the coordinate system (RCS) which rotates around the $z$ axis with frequency $\Omega$, the spins are subjected to the alternating field

$$\vec{H}(t) = \frac{1}{\gamma}\left[\Delta(t)\vec{k} + \omega_1\vec{i}\right],$$

(the non-resonant component of $\vec{H}_1(t)$ field is omitted), where $\Delta(t) = \Delta_0 + \Delta_M \cos\omega t$ is the instant detuning, $\Delta_0 = \omega_0 - \Omega$, $\omega_1 = \gamma H_1$, $\Delta_M = \gamma H_M$ is the modulation amplitude in frequency units, $\vec{k}$ is the $z$ axis unit vector. Let us suppose that the modulation frequency is low enough, so that

$$\omega << T_2^{-1}, \quad T_2 |d\Delta(t)/dt| << T_2^{-1}. \tag{2}$$

The SS is described by the modified Provotorov's equations [1, 6] which in the variables of Eq. (1) read:

$$\frac{dM_z(\tau)}{d\tau} = -\frac{2W(\Delta(\tau))}{\omega}\left\{M_z(\tau) - \frac{\Delta(\tau)}{\omega_d}M_d(\tau)\right\} - \frac{M_z(\tau) - M_0}{\omega T_1},$$

$$\frac{dM_d(\tau)}{d\tau} = \frac{2W(\Delta(\tau))}{\omega} \cdot \frac{\Delta(\tau)}{\omega_d}\left\{M_z(\tau) - \frac{\Delta(\tau)}{\omega_d}M_d(\tau)\right\} - \frac{M_d(\tau) - M_0\omega_d/\omega_0}{\omega T_1'}, \tag{3}$$

where $\tau = \omega t$ is the dimensionless time; $W(x) = (\pi\omega_1^2/2)\cdot\varphi(x)$, $\varphi(x)$ is the resonance line shape, $T_1' = T_1/\alpha$ is the spin-lattice relaxation time of dipole magnetization, $\alpha = 2$ or $3$, depending on the degree of spin correlation [1]. In the relaxation terms of the Eq.(3), the inequality $\omega_0 \gg \Delta_M$ is supposed.

Apart from dilution effects [10] which are of no significance in our consideration, one could assume $T_2^{-1} \sim \omega_d$ and present the second condition of Eq. (2) in the form of $\Delta_M \ll \omega_d^2/\omega$. So, the system Eq. (3) is correct in the range of frequencies and modulation amplitudes defined by the inequalities:

$$\omega \ll \omega_d, \quad \Delta_M \ll \omega_d^2/\omega \text{ or } \omega \ll \omega_d^2/\Delta_M \tag{4}$$

### 3. Qualitative picture of relaxation oscillations

For the system of linear differential equations with periodical coefficients, Eq. (3), the exact solution has not been found so far. The case of small modulation amplitudes has been investigated in details in Ref. [1]. Here we restrict ourselves with the deep modulation, i.e. it is assumed that

$$\omega_d \ll \Delta_M \ll \omega_d^2/\omega \tag{5}$$

in correspondence with Eq. (4).

From the following it will be clear that the most interesting effect of the relaxation oscillations is observed at the frequencies of $\omega \sim T_1^{-1}$. In such a case, the coefficients at the difference of $\left( M_z - \dfrac{\Delta(\tau)}{\omega_d} M_d \right)$ obey the following requirements:

$$\frac{2W(\Delta(\tau))}{\omega} \sim \omega_1^2 T_1 T_2 \equiv S \gg 1, \quad \frac{2W(\Delta(\tau))}{\omega} \cdot \frac{\Delta(\tau)}{\omega_d} \sim S\frac{\Delta_M}{\omega_d} \gg 1, \tag{6}$$

where $S$ is the saturation parameter, and $T_2 = \pi\varphi_2(0)$.

The relations Eq. (6) correspond to the presence of small parameters at upper derivatives; nevertheless, this case needs particular consideration (see, as example, Ref. [11]). During the dimensionless modulation period $\tau_M = \omega T_M = 2\pi$, the condition of the exact resonance $\Delta(\tau) = 0$ holds twice at the time moments $\tau_1$ and $\tau_2 = 2\pi - \tau_1$, which are the roots of the equation $\Delta_0 + \Delta_M \cos\tau = 0$. As far as $\Delta_M \gg \omega_d$, the saturation of resonance takes place twice (in the vicinity of the time moments $\tau_1$ and $\tau_2$) during the time $\tau_M$. The duration $\tau_p$ of saturation experienced by the SS can be determined from the inequality $|\Delta(\tau)| \leq \omega_d$. By the order of magnitude, it equals to $\tau_p \sim 2\pi\omega_d/\Delta_M$. So, according to Eq.(4), one has $\tau_p \gg \omega/\omega_d$.

Let us introduce the time $\tau_s = \omega[2W(0)]^{-1} = \omega(\omega_1^2 T_2)^{-1}$ characterizing the saturation establishment in the SS: during the time interval of the order of $\tau_s$, the Zeeman and dipolar temperatures in the RSC are equalized. In terms of magnetization, this means that in the time about $\tau_s$ after the beginning of saturation, the quasi-equilibrium is established in the SS and so the relation

$$M_z(\tau) = M_d(\tau)\Delta(\tau)/\omega_d \qquad (7)$$

becomes valid. Later on the $M_z(\tau)$ and $M_d(\tau)$ values change adiabatically, i.e., under conditions of Eq. (7), up to the time moment of $\tau_p$, when the saturation is finished.

In this discussion it is supposed that inequality $\tau_p > \tau_s$ takes place. Let us call for a more severe condition $\tau_p >> \tau_s$, i.e., $\omega_d / \Delta_M >> \tau_s$ providing basically the adiabatic character of the saturating process, except for a narrow initial part with the duration of $\tau_s << \tau_p$. On the other hand, this condition means that the time $\tau_s$ of the establishment of quasi-equilibrium in the SS is much shorter that the duration of the resonance line sweeping, $\tau_{sweep} \sim \omega \cdot \omega_d / |\dot{\Delta}| \sim \omega_d / \Delta_M \sim \tau_p$. In order to provide the isolation of SS from the lattice during the saturation process, the condition of $\tau_p << \omega T_1$, i.e., $(\omega T_1)^{-1} << \Delta_M / \omega_d$ must be fulfilled.

The time interval when the free SL relaxation occurs is obviously of the order of $2\pi - 2\tau_p \approx 2\pi$. Let us suppose that $\tau_M \sim \omega T_1$, i.e., $\omega \sim T_1^{-1}$. In this case, the equilibrium state in the SS has enough time to be established during the intervals between the saturation periods.

The combination of all these conditions gives

$$1 \leq (\omega T)^{-1} << \Delta_M / \omega_d << \tau_s^{-1} << \omega_d / \omega . \qquad (8)$$

Inequalities (8) hold at the intermediate saturation of resonance ($T_1^{-1} << W << T_2^{-1}$) for the sufficiently low modulation frequencies ($\omega << \omega_d$) and a deep modulation ($\Delta_M >> \omega_d$). They are optimal for observation of the relaxation oscillations of magnetizations $M_z(\tau)$ and $M_d(\tau)$.

The expected temporal behavior of $M_z(\tau)$ and $M_d(\tau)$ under conditions of Eq.(8) is as follows. During the modulation period $\tau_M$, the SS is subjected to two saturating pulses acting near the time instants $\tau_1$ and $\tau_2$ with the duration $\tau_p << \tau_M$. During the action of the $\tau_1$ pulse, the resonance line is passed from the left side ($\Omega < \omega_0 \to \Omega > \omega_0$), so the dipole subsystem is cooled ($\beta_d >> \beta_L$). In contrast, during the action of the $\tau_2$ pulse one has ($\Omega > \omega_0 \to \Omega < \omega_0$) and the dipole subsystem is heated up to negative temperatures ($\beta_d < 0, |\beta_d| >> \beta_L$). The interval $\tau_2 - \tau_1$ between pulses can be changed by varying the detuning $\Delta_0$. In particular, at $\Delta_0 = 0$ one has $\tau_1 = \pi/2$, $\tau_2 = 3\pi/2$; when $\Delta_0$ is positive and increasing, one has $\tau_1 \to \tau_2 \to \pi$; and, finally, at $\Delta_0 < 0$ an increase of $|\Delta_0|$ leads to $\tau_1 \to 0$, $\tau_2 \to 2\pi$. During the rest of time $2\pi - 2\tau_p \approx 2\pi$, saturation is absent and free S-L relaxation takes place. During the saturation periods, $M_z$ $M_d$ change quickly (with a rate $\sim W$) from their equilibrium values to saturated ones, whereas in the S-L relaxation periods they change slowly (with a rate of $T_1^{-1} \ll W$), and the equilibrium state has enough time to be established. In the established state, as calculations show, the change of magnetization has a periodic character with the periods of $2\pi$ at $\Delta_0 \neq 0$ and of $\pi$ at $\Delta_0 = 0$.

Such interchange of slow and fast (stepwise) evolution caused by dissipation processes is typical of the relaxation oscillations in nonlinear autonomous systems [11, 12]. In our case, these oscillations appear in a linear system with periodic coefficients and are related with a periodic sharp interchange of two modes, the saturation and relaxation ones, which correspond to quite different quasi-equilibrium and equilibrium states.

Various aspects of the influence of the harmonic field modulation on spin systems obeying the Bloch equations were discussed both theoretically and experimentally in Ref. [13]. In particular, the stepwise oscillations of $M_z(\tau)$ were observed. Some other modulation effects, such as the quasi-adiabatic single-shot passage through the resonance, were investigated on the basis of Provotorov's equations in Refs. [1, 14].

**4. Quantitative theory of relaxation oscillations**

As it follows from the above presented considerations and what is supported by the following calculations, for the sufficiently narrow resonance line shape, the picture of relaxation oscillations is not sensitive very much to its details. For this reason, to simplify calculations, we will approximate $\varphi(x)$ by rectangular form with the half-width of $T_2^{-1} \sim \omega_d$:

$$\varphi(x) = \begin{cases} T_2/\pi, & |x| < T_2^{-1} \\ 0, & |x| > T_2^{-1} \end{cases}. \quad (9)$$

Correspondingly, the probability dependence $2W(\Delta(\tau))$ on $\tau$ is expressed as in Fig. 1.

One may call the $\tau_i^\pm = \tau_i \pm \tau_p/2$ $(i = 1, 2)$ moments as the borders of saturation inversion. This definition is, obviously, not exact, but the final results are insensitive to the choice of $\tau_i^\pm$ with the accuracy up to $\tau_p(\omega T_1^{-1})^{-1} \ll 1$

In the relaxation mode ($W = 0$), the solution of Eq. (3) with the initial values $M_z(0)$ and $M_d(0)$ could be expressed as:

$$M_z(\tau) = M_0 + [M_z(0) - M_0]\exp[-\tau(\omega T_1)^{-1}],$$

$$M_d(\tau) = M_0\omega_d/\omega_0 + [M_d(0) - M_0\omega_d/\omega_0]exp[-\tau(\omega T_1')^{-1}] \qquad (10)$$

Studying the saturation mode ($T_1^{-1} = 0$) is more complicated. In this case, the SS is described by the equations:

$$\frac{dM_z(\tau)}{d\tau} = -\frac{2W(\Delta(\tau))}{\omega}\left[M_z(\tau) - \frac{\Delta(\tau)}{\omega_d}M_d(\tau)\right],$$

$$\frac{dM_d(\tau)}{d\tau} = \frac{\Delta(\tau)}{\omega_d} \cdot \frac{2W(\Delta(\tau))}{\omega}\left[M_z(\tau) - \frac{\Delta(\tau)}{\omega_d}M_d(\tau)\right], \qquad (11)$$

Eq. (11) is valid into the very narrow regions $\tau_i^- \leq \tau \leq \tau_i^+$ with the width of $\tau_p \ll 2\pi$.

Let us introduce the unknown functions:

$$V(\tau) = \left[M_z(\tau) - \frac{\Delta(\tau)}{\omega_d}M_d(\tau)\right] \cdot \left[(1 + \Delta^2(0)/\omega_d^2)/(1 + \Delta^2(\tau)/\omega_d^2)\right]^{1/2},$$

$$U(\tau) = \left[\frac{\Delta(\tau)}{\omega_d} \cdot M_z(\tau) + M_d(\tau)\right] \cdot \left[(1 + \Delta^2(0)/\omega_d^2)/(1 + \Delta^2(\tau)/\omega_d^2)\right]^{1/2}. \qquad (12)$$

In the $U$, $V$ variables the system (11) takes the form:

$$\frac{dV(\tau)}{d\tau} = -V(\tau)[\tau_R(\Delta(\tau))]^{-1} + v(\tau)U(\tau), \qquad \frac{dU(\tau)}{d\tau} = -v(\tau)V(\tau), \qquad (13)$$

where

$$\tau_R^{-1}(\Delta(\tau)) = \frac{2W(\Delta(\tau))}{\omega} \cdot [1 + \Delta^2(\tau)/\omega_d^2], \quad v(\tau) = \Delta_M \omega_d[\Delta^2(\tau) + \omega_d^2]^{-1} \sin\tau. \qquad (14)$$

The use of $U$ and $V$ variables is convenient because the following estimations are valid:

$$\tau_R^{-1}(\Delta(\tau)) \sim \frac{\Delta_M^2 T_2^2}{\tau_s} \gg 1, \qquad |v(\tau)| \sim \frac{\omega_d}{\Delta_M}|\sin\tau| \ll 1,$$

due to the inequalities $\tau_s^{-1} \gg \Delta_M T_2 \gg 1$, see Eq. (8).

It follows from Eq. (14) that $V(\tau)$ changes quickly and $U(\tau)$ changes slowly. This enables one to find an approximate solution of Eq. (13).

It is evident that $V(\tau) \to 0$, i.e. $M_z \to M_d\Delta(\tau)/\omega_d$ exponentially with the characteristic time $\sim \tau_s$. This means, as it has been already noted earlier, that the adiabaticity is established for the time $\sim \tau_s$ after the saturation is "turned on". The further change of $M_z$ and $M_d$ during the duration of $\sim \tau_p - \tau_s \approx \tau_p$ takes place in the "lock-in" mode, with obeying of the condition Eq. (7).

As it follows from Eq. (13), $V(\tau) \to 0$, $U(\tau) \cong const$ at $\tau_R^{-1} \to \infty$. Thus, for a sufficiently large $\tau_R^{-1}$ values, the approximate solution of Eq. (13) is as follows:

$$V^{(i)}(\tau) = V(\tau_i^-) exp\left[-\int_{\tau_i^-}^{\tau} d\tau'/\tau_R(\Delta(\tau'))\right], \quad U^{(i)}(\tau) = U(\tau_i^-) = U(\tau_i^+) = const, \quad (15)$$

where $i = 1, 2$ number the regions of saturation near the moments of $\tau_1$ and $\tau_2$, and $V(\tau_i^-)$, $U(\tau_i^-)$ could be expressed through $M_{zi}^- \equiv M_z(\tau_i^-)$, $M_{di}^- \equiv M_d(\tau_i^-)$ assuming $\tau = \tau_i^-$ in Eq. (12).

Introducing Eq.(15) in Eq.(12), one could write the solution of Eq. (13) in the form:

$$M_{zi}(\tau) = M_{zi,ad}(\tau) + [M_{zi}^- - M_{di}^- \Delta(\tau_i^-)/\omega_d] R_i(\tau), \quad (16)$$

$$M_{di}(\tau) = M_{di,ad}(\tau) - \Delta(\tau)/\omega_d [M_{zi}^- - M_{di}^- \Delta(\tau_i^-)/\omega_d] R_i(\tau),$$

where

$$M_{zi,ad}(\tau) = \Delta(\tau)/\omega_d \cdot [M_{di}^- + M_{zi}^- \Delta(\tau_i^-)] \cdot \{[1 + \Delta^2(\tau_i^-)/\omega_d^2] \cdot [1 + \Delta^2(\tau)/\omega_d^2]\}^{-1/2} \quad (17)$$

$$M_{di,ad}(\tau) = [M_{di}^- + M_{zi}^- \cdot \Delta(\tau_i^-)/\omega_d] \cdot \{[1 + \Delta^2(\tau_i^-)] \cdot [1 + \Delta^2(\tau)/\omega_d^2]\}^{-1/2}$$

and the function $R_i(\tau)$ is given by the following expression:

$$R_i(\tau) = \{[(1 + \Delta^2(\tau_i^-)/\omega_d^2) \cdot (1 + \Delta^2(\tau)/\omega_d^2)]^{-1/2} \exp\left\{-\int_{\tau_i^-}^{\tau} d\tau' \tau_R^{-1}(\Delta(\tau'))\right\}. \quad (18)$$

It is evident that $M_{zi,ad}$ and $M_{di,ad}$ are related to the adiabatic condition, Eq. (7). To facilitate the analysis of the obtained solutions, let us evaluate the exponent in Eq. (18). Accordingly to Eq. (14), one has

$$I \equiv \int_{\tau_i^-}^{\tau} d\tau' \tau_R^{-1}(\Delta(\tau')) = \int_{\tau_i^-}^{\tau} d\tau' \frac{2W(\Delta(\tau'))}{\omega} [1 + \Delta^2(\tau')/\omega_d^2].$$

The probability $W$ differs from zero in very narrow ranges $\sim \tau_p \ll 2\pi$ near the moments $\tau_1$ and $\tau_2$, therefore one could assume in Eq.(18):

$$2W(\Delta(\tau'))/\omega \cong 2W(0)/\omega = \tau_s^{-1},$$

resulting in

$$I \cong \frac{1}{\tau_S} \int_{\tau_i^-}^{\tau} d\tau' [1 + \Delta^2(\tau')/\omega_d^2].$$

Using the known theorem on the average for the definite integral, one gets finally:

$$I \cong (\tau - \tau_i^-)/\bar{\tau}_R, \quad \bar{\tau}_R = \tau_S /[1 + \Delta^2(\bar{\tau})/\omega_d^2] < \tau_S, \quad (19)$$

where $\tau_i^- < \bar{\tau} < \tau$.

So, the $R_i(\tau)$ function reads:

$$R_i(\tau) = \{[1 + \Delta^2(\tau_i^-)/\omega_d^2] \cdot [1 + \Delta^2(\tau)/\omega_d^2]\}^{-1/2} \exp[-(\tau - \tau_i^-)/\bar{\tau}_R]. \quad (20)$$

It is easily seen that $R_i(\tau)$ tends to zero exponentially at $|\tau - \tau_i^-| > \tau_s$. Thus, the $M_{zi,ad}$ and $M_{di,ad}$ values, which are related by Eq. (7), correspond to the adiabatic range. This regime is quickly (exponentially) established for the small time $\sim \tau_s \ll \tau_p$ after the saturation is "turned on" and retained up to the moment $\tau_i^+$ when the saturation is "turned off". Since $\tau_p \ll 2\pi$, one can perform the expansion of the instant detuning $\Delta(\tau)$ in the vicinity of the moment $\tau = \tau_i$ during the saturation ranges:

$$\Delta(\tau) = (-1)^i \kappa \omega_d X_i(\tau), \tag{21}$$

where

$$\kappa = \Delta(\tau_1^-)/\omega_d = \pi\sqrt{\Delta_M^2 - \Delta_0^2}/\Delta_M \sim \pi, \quad X_i(\tau) = 2(\tau - \tau_i)/\tau_p \in [-1,1]. \tag{22}$$

As a result, the adiabatic values of susceptibilities, Eq. (17), take the form:

$$M_{zi,ad}(\tau) = (-1)^i \kappa X_i(\tau)\left[M_{di}^- + (-1)^{i+1}\kappa M_{zi}^-\right] \cdot \left\{(1+\kappa^2)[1+\kappa^2 X_i^2(\tau)]\right\}^{-1/2}.$$

$$M_{di,ad}(\tau) = \left[M_{di}^- + (-1)^{i+1}\kappa M_{zi}^-\right] \cdot \left\{(1+\kappa^2)[1+\kappa^2 X_i^2(\tau)]\right\}^{-1/2}, \tag{23}$$

and the function of Eq. (20) is defined by the expression:

$$R_i(\tau) = \left\{(1+\kappa^2)[1+\kappa^2 X_i^2(\tau)]\right\}^{-1/2} \exp\{-(\tau - \tau_i^-)/\bar{\tau}_R\}. \tag{24}$$

So, in the adiabatic region, the passing through the resonance is nearly linear, though the field modulation is cosine.

Using Eqs. (10), (16), (21), (23) and (24), one could write the solutions for $M_{zk}(\tau)$, $M_{dk}(\tau)$ (k=1, 2, 3, 4, 5) in the regions of $1 \equiv [0, \tau_1^-]$, $2 \equiv [\tau_1^-, \tau_1^+]$, $3 \equiv [\tau_1^+, \tau_2^-]$, $4 \equiv [\tau_2^-, \tau_2^+]$ and $5 \equiv [\tau_2^+, 2\pi]$. Carrying out the concatenation of solutions in points $\tau_1^\pm, \tau_2^\pm$ and introducing conditions $M_{z5}(2\pi) = M_z(0)$, $M_{d5}(2\pi) = M_d(0)$ at $\Delta_0 \neq 0$, or conditions $M_{z3}(\pi) = M_z(0)$, $M_{d3}(\pi) = M_d(0)$ at $\Delta_0 = 0$, providing the periodicity of functions $M_z(\tau)$, $M_d(\tau)$, one could find the solution of Eq. (3) in the explicit form. The obtained expressions are rather cumbersome, so some simplifying assumptions are accepted. We suppose that the equilibrium is practically established during the spin-relaxation periods. This means that

$$M_z(\tau_{1,2}^-) \cong M_0, \quad M_d(\tau_{1,2}^-) \cong M_0 \omega_d/\omega_0. \tag{25}$$

Let us discuss the obtained results. The calculated functions $M_z(\tau)$, $M_d(\tau)$ for the case of exact resonance $\Delta_0 = 0$, when $\tau_1 = \pi/2$, $\tau_2 = 3\pi/2$ and the condition Eq. (25) is obeyed, are plotted in Fig. 2.

In the region of $\tau_i$-saturation ($i = 1, 2$) there is a narrow border layer $\tau_i^- \leq \tau \leq \tau_i^- + \tau_S$ where the resonance detuning $\Delta(\tau) \approx \Delta(\tau_i^-) = const$ and $M_z(\tau)$ changes quickly (exponentially) and non-adiabatically from the value of $M_z(\tau_i^-) \approx M_0$ up to the value of $M_z(\tau_1^- + \tau_S) \approx M_0 \kappa^2 /(1 + \kappa^2)$, in correspondence with Eq. (23). Then the adiabatic, close to linear inversion of the Zeeman magnetization takes place:

$$M_z(\tau_i^- + \tau_S) \to M_z(\tau_i^+) = -M_z(\tau_i^- + \tau_S),$$

which is caused by the inversion at $\Delta(\tau_i^+) = -\Delta(\tau_i^-)$. It is evident that $M_z(\tau_i) = 0$.

The $M_d(\tau)$ behavior is clear from the following. During the first interval of non-adiabaticity, one has $\Delta(\tau) \approx \Delta(\tau_1^-) = \kappa \omega_d > 0$ ($\omega_0 > \Omega$), and the dipole subsystem "restores" the energy of the Zeeman subsystem. As a result, it is cooled down to

$$\beta_d^{-1}(\tau_1^- + \tau_s) = \frac{1 + \kappa^2}{\kappa} \cdot \frac{\omega_d}{\omega_0} \beta_L^{-1} \gg \beta_L^{-1}. \tag{26}$$

During the second interval of non-adiabaticity $\Delta(\tau) \approx \Delta(\tau_2^-) = -\kappa \omega_d < 0$ ($\omega_0 < \Omega$), and the dipole system "receives" the energy from the Zeeman subsystem being heated up to the negative temperature

$$\beta_d^{-1}(\tau_2^- + \tau_s) = -\frac{1 + \kappa^2}{\kappa} \cdot \frac{\omega_d}{\omega_0} \beta_L^{-1}, \quad \left|\beta_d^{-1}\right| \gg \beta_L^{-1}.$$

In the adiabatic periods, the $|M_d|$ value changes insignificantly from $|M_d|_{min} = |M_d(\tau_i^- + \tau_S)| = M_0 \kappa (1 + \kappa^2)^{-1}$ up to $|M_d|_{max} = |M_d(\tau_i)| = (1 + \kappa^2)^{1/2} |M_d|_{min}$ and further from $|M_d|_{max}$ to $|M_d|_{min}$ at the moment $\tau = \tau_i^+$ of the saturation termination. Between the saturations, during the time - $\tau_2^- - \tau_1^+ \equiv \Delta \tau_L$, the recovery of the equilibrium magnetizations takes place. By changing the $\Delta_0$ value, one is able either to enhance the interval $\Delta \tau_L$ (at $\Delta_0 < 0$, $|\Delta_0|$ increases), or to reduce this interval (at $\Delta_0 > 0$, $\Delta_0$ increases), and thereby to influence the values of $M_{z,d}(\tau_i^+)$ which are reached with the participation of spin-lattice relaxation. This opens the possibility of the steady-state measurement of the spin-lattice relaxation times $T_1$ and $T_1' = T_1 / \alpha$. The evolution of $M_z(\tau)$ and $M_d(\tau)$ is repeated with the modulation period at $\Delta_0 \neq 0$ and with the half-period of modulation at $\Delta_0 = 0$.

If the dipole subsystem is neglected ($M_d = 0$), the picture of the relaxation oscillations $M_z(\tau)$ can be obtained by the formal transition $\kappa \to 0$ (i.e., $\omega_d \to \infty$). In this case, in the $\tau_i$-saturation region, the magnetization $M_z(\tau)$ changes exponentially quickly from $M_0$ down to zero for a small time $\tau_S$. It remains equal to zero up to the end of saturation (the linear part, existing at $M_d \neq 0$, is absent), and then the recovery of the equilibrium value of $M_0$ begins to occur. The inversion of $M_z(\tau)$ does not take place. (In Ref. [13], the possibility of the $M_z(\tau)$ inversion in strong fields ($\omega_1 \gg \omega_d$) is considered; this case is beyond the scope of our work).

Thus, the $M_z(\tau)$ inversion under conditions of the deep modulation and intermediate saturation ($\omega_1 \ll \omega_d$) could serve as experimental criterion for estimation of the dipole-temperature effect. As it is seen from the $M_z(\tau)$ plot (Fig. 2), the maximum inverted magnetization amounts to $\sim M_0$. Apparently, this enables one to employ this effect in two-level masers. Note that the maximum value of the alternative component of the longitudinal magnetization in V.A. Atsarkin's effect [4, 5], where the modulation is small, is of the order of $|\Delta M_z(\tau)|_{max} \sim M_0 \Delta_M / \omega_d \ll M_0$.

The performed consideration shows that, provided the conditions Eq. (8) are fulfilled, the continuous modulation of magnetic field is equivalent to the effect of two short saturation pulses which act during each modulation period., The repetition order and duration of these pulses can be adjusted with the help of the resonance detuning $\Delta_0$ and modulation amplitude $H_M$. The response on these pulses is the appearance of the relaxation oscillations, their form being varied in a wide range by changing the modulation frequency $\omega \ll T_2^{-1}$ and $\Delta_0$. Thus, the SS could be also used as the relaxation generator for the producing of magnetization pulses of different shapes.

## 5. Peculiarities of the lock-in detection signals in the relaxation oscillations mode

Let us discuss now peculiarities of the absorption $V$ and dispersion $U$ signals registered with the transverse RF magnetic field ($\Omega, 2H_1$) in the mode of relaxation oscillations.

By definition [1],

$$U(t) = Sp\{\rho^*(t)\hat{M}_x\}, \quad V(t) = Sp\{\rho^*(t)\hat{M}_y\}, \tag{27}$$

where $\rho^*(t)$ is the statistical operator in the RCS, $\hat{M}_{x,y} = -\hbar\gamma S_{x,y}$ are the magnetic momentum operators.

So, $U$ and $V$ are projections of the magnetization vector on the transverse axes of RCS. It could be shown [1, 6], that in the low-frequency region ($\omega \ll T_2^{-1}$), the following expressions hold in the $M_z(\tau), M_d(\tau)$ variables:

$$U(\tau) = \omega_1/\omega_d \cdot \{M_d(\tau) - \omega_d J(\Delta(\tau))[M_z(\tau) - M_d(\tau)\Delta(\tau)/\omega_d]\}$$

$$V(\tau) = -\pi\omega_1 f(\Delta(\tau))[M_z(\tau) - M_d(\tau)\Delta(\tau)/\omega_d], \qquad (28)$$

where $J(x) = \int_{-\infty}^{\infty} \frac{f(y)}{y-x} dy = -J(-x)$.

The terms of $U$ and $V$ signals are related with the fact that at the absence of modulation ($\Delta_M = 0$)

$$U = 2H_1\chi'(\Omega), \ V = -2H_1\chi''(\Omega),$$

where $\chi'$ и $\chi''$ are the real and imaginary parts of the complex susceptibility in B.N. Provotorov's theory [1, 8]. As it is well known [1], in the lock-in detection method the harmonics of periodical functions $U(\tau)$ and $V(\tau)$ are measured on frequencies $k\omega$ ($k = 1,2,3,\ldots$). To find them out, let us decompose $U(\tau)$ and $V(\tau)$ in the Fourier series:

$$(U,V)(\tau) = \sum_{k=-\infty}^{\infty}(U,V)_k \exp(ik\tau) = (U,V)_0^C/2 + \sum_{k=1}^{\infty}\{(U,V)_k^C \cos k\tau + (U,V)_k^S \sin k\tau\}.$$

Experimentally, the real values of $(U,V)_k^{(C,S)}$ are directly measured. They are related with the complex Fourier coefficients $U_k = U_{-k}^*$, $V_k = V_{-k}^*$ by the expressions:

$$(U,V)_k^{(C)} = (U,V)_k + (U,V)_k^*, \ (U,V)_k^{(S)} = i[(U,V)_k - (U,V)_k^*].$$

Let us firstly calculate $V_k$. It is evident that

$$V_k = (2\pi)^{-1} \int_0^{2\pi} d\tau V(\tau) \exp(-ik\tau).$$

Using Eq. (28) and taking into account that the function $f(\Delta(\tau))$ differs from zero only into the saturation intervals, it can be easily shown that in the relaxation oscillations mode $V_k \cong 0$, i.e., the lock-in absorption signal is absent.

Now it is evident that the term with $J(\Delta(\tau))$ in Eq. (28) does not contribute into $U_k$. Therefore

$$U(\tau) \cong M_d(\tau)\omega_1/\omega_d, \ U_k = \omega_1(2\pi\omega_d)^{-1}\int_0^{2\pi} d\tau \, M_d(\tau)\exp(-ik\tau). \qquad (29)$$

It is not difficult to show that, for $k\tau_p \ll 1$, the saturation regions in Eq. (29) can be neglected, because

$$\int_{\tau_1^-}^{\tau_1^+} d\tau \, M_d(\tau) + \int_{\tau_2^-}^{\tau_2^+} d\tau \, M_d(\tau) = 0.$$

Hence,

$$U_k = \omega_1(2\pi\omega_d)^{-1}\left\{\int_{\tau_1^+}^{\tau_2^-} d\tau \, M_{d3}(\tau)\exp(-ik\tau) + \int_{\tau_2^+}^{2\pi} d\tau \, M_{d5}(\tau)\exp(-ik\tau)\right\}. \qquad (30)$$

Using Eq. (10) and allowing for the concatenation in the $\tau_1^+$ and $\tau_2^+$ moments, the solutions in regions 3 and 5 can be easily obtained:

$$M_{d3}(\tau) = M_0 \omega_d / \omega_0 + \left[M_{d2}(\tau_1^+) - M_0 \omega_d / \omega_0\right] \cdot \exp\left(-\frac{\tau - \tau_1^+}{\omega T_1'}\right),$$

$$M_{d5}(\tau) = M_0 \omega_d / \omega_0 + \left[M_{d4}(\tau_2^+) - M_0 \omega_d / \omega_0\right] \cdot \exp\left(-\frac{\tau - \tau_2^+}{\omega T_1'}\right), \tag{31}$$

where $M_{d3}(\tau_1^+)$, $M_{d5}(\tau_2^+)$ are the values of $M_d(\tau)$ at the moments of saturation finishing.

Let us consider the detuning regions $|\Delta_0| \ll \Delta_M$ and assume, for the simplicity sake, that $\exp\left(-\frac{\pi}{\omega T_1}\right) \approx \exp\left(-\frac{\pi}{\omega T_1'}\right) \approx 0$. Then

$$\tau_1 \approx \pi/2, \quad \tau_2 \approx 3\pi/2, \quad M_{d2}(\tau_1^+) = -M_{d4}(\tau_2^+) = M_0 \kappa (1 + \kappa^2). \tag{32}$$

Substituting Eqs. (31), (32) in Eq. (30) and performing elementary calculations, one gets the following expressions for the odd harmonics:

$$U_{2k+1} = M \frac{\omega_1}{\pi \omega_d} \frac{\kappa}{1 + \kappa^2} \frac{i(-1)^{k+1}}{(2k+1)i + \frac{1}{\omega T_1'}}.$$

The even harmonics $U_{2k}$ appear to be proportional to the small value of $M_0 \frac{\omega_d}{\omega_0} \ll M_0$ and can be neglected.

The values of $U_{2k+1}^{(C)}$ and $U_{2k+1}^{(S)}$ measured in the experiment become equal to:

$$U_{2k+1}^{(C)} = M_0 \frac{2\kappa}{1 + \kappa^2} \frac{\omega_1}{\pi \omega_d} \frac{(2k+1)(-1)^{k+1^2}(T_1')^2}{1 + (2k+1)^2 \omega^2 (T_1')^2} = -(2k+1)\omega T_1' U_{2k+1}^{(S)}.$$

In particular, for the first harmonics $(k = 0)$ one gets:

$$U_1^{(C)} = -M_0 \frac{2\kappa}{1 + \kappa^2} \frac{\omega_1}{\pi \omega_d} \cdot \frac{\omega^2 (T_1')^2}{1 + \omega^2 (T_1')^2} = -\omega T_1' U_1^{(S)}.$$

Accordingly to Eq. (26), the values of $U_1^{(C)}, U_1^{(S)} \sim \beta_d$, therefore the measurement of the dispersion signals $U_{2k+1}^{(C)}$ and $U_{2k+1}^{(S)}$ in the relaxation oscillations mode can be used as a "probe" for the determination of the temperature and other parameters ($T_1' = T_1/\alpha$, $\omega_d$) characterizing the dipole subsystem.

FIGURE CAPTIONS

Fig. 1. Time dependence of the probability of the induced resonance transitions under the conditions of deep modulation.

Fig. 2. Time dependence of Zeeman and dipole magnetizations under the conditions of deep modulation.

# FIGURES

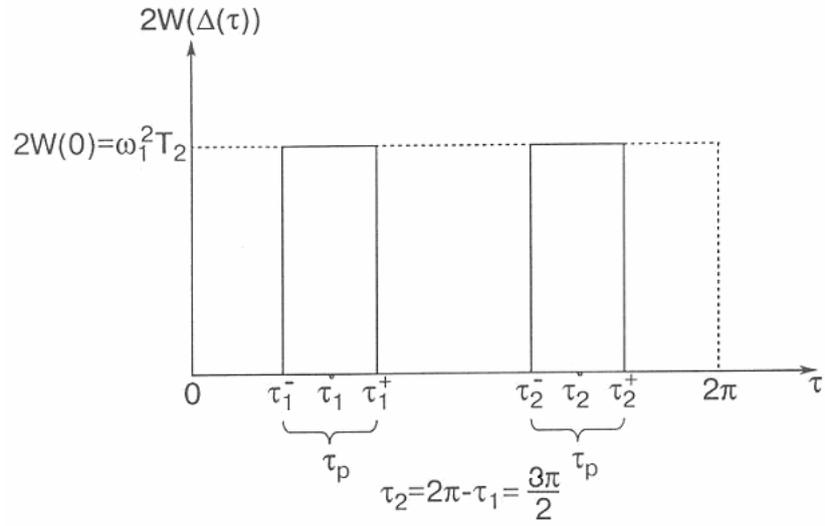

Fig. 1.

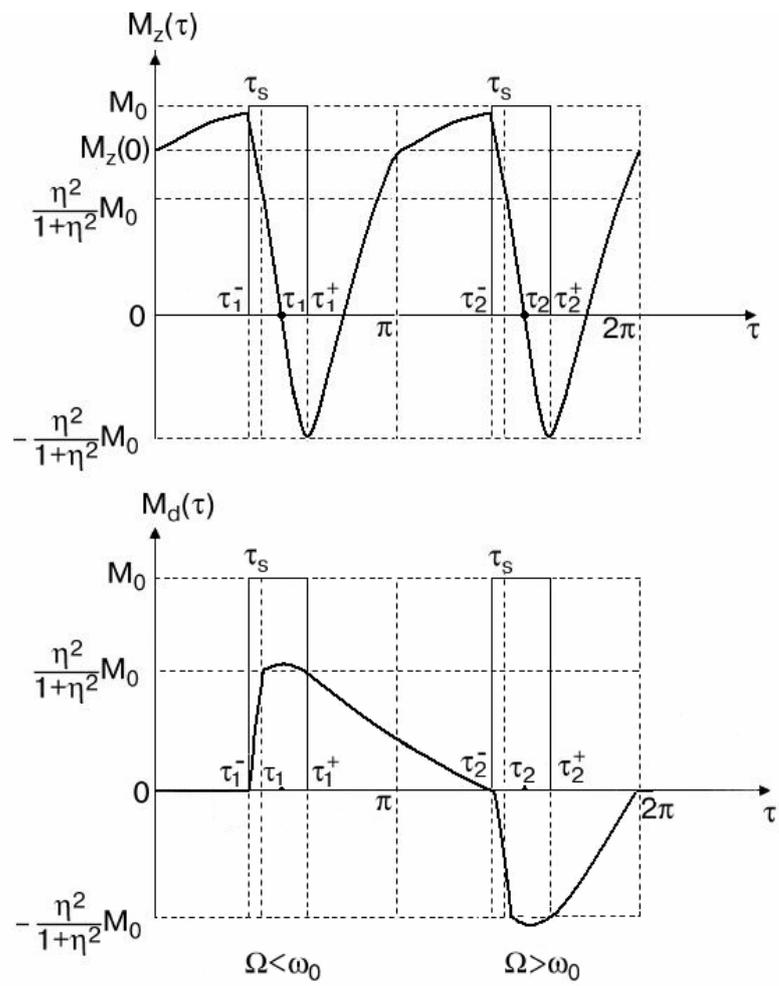

Fig. 2.